\documentclass[conference]{IEEEtran}
\IEEEoverridecommandlockouts
\usepackage{amsmath,stfloats}
\usepackage{graphicx}
\include{psfig}
\usepackage{epsfig}
\usepackage{cite}
\usepackage{array}
\usepackage{psfrag}
\usepackage{amssymb}
\usepackage{mathdots}
\include{psfig}
\usepackage{color}
\usepackage{pstricks,pst-node,pst-text,pst-3d,pst-plot}
\usepackage{psfrag}
\newcounter{mytempeqncnt}
\begin{document}

\title{A New Achievable DoF Region for the 3-user $M\times N$ Symmetric Interference Channel}


\author{\IEEEauthorblockN{Mohamed Amir, Amr El-Keyi, and  Mohammed
Nafie}
\IEEEauthorblockA{Wireless Intelligent Networks Center (WINC) \\
Nile University, Cairo, Egypt.\\ Email: {\tt
mohamed.khalil@nileu.edu.eg, aelkeyi, mnafie@nileuniversity.edu.eg}}
\thanks{This work was supported in part by a grant from the Egyptian National Telecommunications Regulatory
Authority (NTRA)}}

\maketitle

\vspace{-1mm}
\begin{abstract}
In this paper, the $3$-user multiple-input multiple-output
Gaussian interference channel with $M$ antennas at each
transmitter and $N$ antennas at each receiver is considered. It is
assumed that the channel coefficients are constant and known to
all transmitters and receivers. A novel scheme is presented that
spans a new achievable degrees of freedom region. For some values
of $M$ and $N$, the proposed scheme achieve higher number of DoF
than are currently achievable, while for other values it meets the
best known upperbound. Simulation results are presented showing
the superior performance of the proposed schemes to earlier
approaches.
\end{abstract}
\IEEEpeerreviewmaketitle

\section{Introduction}
The capacity of the interference channels has kept Information
theorists busy for more than three decades, e.g.,
\cite{IntChCar},\cite{Bresler_Capacity}. Their work has brought
great understanding to interference channels, but full capacity
characterization has not been found to the moment. In the absence
of precise capacity characterizations, researchers have pursued
asymptotic and/or approximate capacity characterizations. One of
the most important capacity characterizations is the degrees of
freedom (DoF) of the network
\cite{Host}-\nocite{Maddah_Signaling_mimo,Maddah_Signalling,DoFmimoICjafar,Jafar_mimo_X,Jafar_X_Networks,Cadambe_spatial,Maddah_IA_dec,Khandani_KuserMxN,Jafar_MN}\cite{Jafar_cmplex}.
The DoF of wireless interference networks represent the rate of
growth of network capacity with the log of the signal to noise
ratio. Fortunately the number of DoF of an interference channel is
equal to the number of interference-free signaling dimensions in
the network.

The DoF of different forms of interference channels have recently
been found, most of these results depend on interference
alignment. Interference alignment refers to the overlapping of
multiple interfering signals from different transmitters into a
small subspace at each receiver so that the number of interference
free dimensions remaining for the desired signal can be maximized.
The idea of interference alignment first appeared in
\cite{Maddah_Signaling_mimo,Maddah_Signalling} where it was used
with dirty paper coding and successive decoding for proposing an
achievable scheme for the X-network. Interference alignment was
then independently used in \cite{Jafar_mimo_X},
\cite{Maddah_IA_dec} by Sayed Jafar and Maddah-Ali respectively in
studying the 2-user X network. Interference alignment has since
been applied to a various types of networks; it was used by Cadambe and
Sayed Jafar in finding the DoF of the $K$ user interference
channel with time varying channel coefficients \cite{Cadambe_spatial},
\cite{Jafar_MN}, and in finding the DoF of X networks with
of $M$
single antenna transmitters and $N$ single antenna
receivers \cite{Jafar_X_Networks}. Motahari et. al
settled the problem in general in
\cite{Real_Khandani},\cite{forming} with their new form of
interference alignment which used results from Diophantine
approximation in Number theory \cite{OX_NO_Theory} to show that
interference can be aligned based on the properties of rationals
and irrationals. They showed that almost all $K$ user real
Gaussian interference channel fading with constant coefficients
have ${K}/{2}$ DoF \cite{Real_Khandani}. They extended these
results to the $K$ user multiple input multiple output (MIMO)
interference channel when the number of transmit antennas is equal
to the number of receive antennas. They showed that the total
number of DoF is equal to ${KM}/ {2}$ whether the channel has
constant or time varying coefficients.

The $K$ user $M\times N$ Gaussian interference channel was first
studied in \cite{Jafar_MN} where it was shown that the total
number of DoF of the $K$ user $M\times N$ time-varying MIMO
Gaussian interference channel is equal to $K\text{min}(M,N)$ if
$K\leq R$ and $\text{min}(M,N) K {R}/{(R+1)}$ when
$R={\max(M,N)}/{\min(M,N)}$ is equal to an integer. Also, it was
shown that for the $G+2$ user MIMO Gaussian interference channel
where each transmitter has $M$ antennas and each receiver has $GM$
antennas with constant channel coefficients, $GM+
\left({GM}/{\left\lfloor G^2+2G-1\right\rfloor}\right)$ DoF can be
achieved without channel extension. Ghasemi et. al
\cite{Khandani_KuserMxN} presented an achievable scheme that can
achieve $K{MN}/{(M+N)}$ for constant channel with real
coefficients. They showed that their scheme coincides with the new
upperbound on the total number of DoF they found for $K >
{(M+N)}/{\text{gcd}(M,N)}$. It is worth noting that their scheme
assumes no cooperation among transmit and/or receive antennas of
each user. The previous results makes the DoF of the $K$ user
$M\times N$ channel with complex constant coefficients almost an
open problem, and for the channel with real constant coefficients,
the results leave a gap when $ K < {(M+N)}/{\text{gcd}(M,N)}$ for
which the number of DoF is unresolved. The gap of unknown DoF of
different $(M,N)$ pairs is actually wide when the number of users
is small. In this paper we provide achievable scheme for different
cases along this gap. Our scheme is optimal for $M/N \geq
{5}/{3}$. It also achieves the beamforming upperbound for $M/N =
(2L+3)/(2L+1)$ where $L=1,2, \ldots$. For other values of $M/N$,
the proposed schemes do not coincide with the best known
upperbound, nevertheless, they achieves more DoF than previously
known.

\section{3-User $MXN$ Symmetric Interference Channel}

\subsection{System model}
We consider a $3$ user MIMO interference channel where each
transmitter and each receiver is equipped with $M$ and $N$
antennas, respectively. Let $\boldsymbol{V}_i$ denote the $M
\times d_i$ precoding matrix of the $i$th user where $d_i$ is the
number of streams, DoF, transmitted by the $i$th user. We can
write the $N$-dimensional received signal at the $i$th receiver at
the $n$th time instant as \vspace{-2mm}
\begin{equation}\label{Received_signal}
\boldsymbol{y}_{i}(n)=\sum^3_{j=1} \boldsymbol{H}_{i,j}
\boldsymbol{V}_j \boldsymbol{x}_j(n)+ \boldsymbol{z}_i(n) \qquad
i=1,\ldots, 3 \vspace{-2mm}
\end{equation}
where $\boldsymbol{H}_{i,j}$ is the $N \times M$ matrix containing
the channel coefficients from transmitter $j$ to receiver $i$,
$\boldsymbol{z}_i(n)$ is the $N\times 1$ additive white Gaussian
noise vector at the $i$th receiver, $\boldsymbol{x}_j(n)$ is the
$d_j \times 1$ vector of Gaussian coded symbols. The $l$th element
of the vector $\boldsymbol{x}_j(n)$ represents the $l$th stream of
the $j$th transmitter, which is transmit beamformed by the $l$th
column of the matrix $\boldsymbol{V}_j$. The received signal
vector at the $i$th receiver is linearly processed by the $N
\times d_i$ post-processing matrix $\boldsymbol{U}_i$ to extract
the $d_i$ streams sent by the $i$th transmitter.
\vspace{-4mm}
\subsection{Upperbounds on the DOF}
\vspace{-1mm}
\subsubsection{General Upperbound}

Following the footsteps of the proof of the upperbound on the DoF
of the $K$ user $M \times N$ Interference Channel DoF in
\cite{Khandani_KuserMxN}, the total number DoF is bounded above by
\vspace{-2mm}
\begin{equation} \label{upperbound1}
\sum_{i=1}^3 d_i\leq
\text{min}\left\{3M,3N,\text{max}\{2M,N\},\text{max}\{2N,M\}\right\}
\vspace{-2mm}
\end{equation}
where the above upperbound can be derived by grouping two
transmitters into one transmitter with $2M$ antennas and grouping
the corresponding two receivers into a single receiver with $2N$
antennas. The upperbound in (\ref{upperbound1}) is derived from
the resulting two-user interference channel using the results in
\cite{DoFmimoICjafar}. Note that the DoF of a 3-user $M \times M$
interference channel is known to be $(3/2) M$.  Since adding
antennas to the transmitters/receivers cannot decrease the DoF, we
can add more antennas at the transmitters or receivers to convert
the system into an interference channel with $\max\{M,N\}$
antennas at each transmitter and receiver. Hence, the total number
DoF is also bounded above by $({3}/{2})\text{max}\{M,N\}$. Hence,
we can augment the upperbound in (\ref{upperbound1}) as,
\vspace{-4mm}
\normalsize

\footnotesize
\begin{equation}
\sum_{i=1}^3 d_i
\!\leq\!\text{min}\!\left\{\!3M,3N,\text{max}\{2M,N\},\text{max}\{M,2N\},
\frac{3}{2}\text{max}\{M,N\}\!\right\}\label{upperbound1_1}
\end{equation}
 \normalsize

\subsubsection{Beamforming Upperbound}
In \cite{Yetis_Feasibility}, it was shown that the total number
DoF of a $3$-user symmetric $M \times N$ Interference Channel that
can be achieved using beamforming only is bounded above by
\begin{equation}\label{upperbound2}
\sum_{i=1}^3 d_i \leq \frac{3}{4} (M+N).
\end{equation}

\subsection{Achievable Schemes}
We assume-without loss of generality-that $M$ is larger than $N$.
Note that in the case of $M < N$, the precoding/decoding matrices
of the transmitters/receivers can be obtained from the
decoding/precoding matrices of the 3-user $N \times M$ reciprocal
channel because of the the reciprocity of the problem. Note that
typically for $M > N$, classical interference alignment is not
enough for achieving the DoF of this network as it achieves at
most ${N}/{2}$ DoF per user.

In the proposed schemes, the nullspaces of the channels to the
non intended receivers are used to mitigate the effect of
interference. We show below that for $M \geq
{5N}/{3}$, there are achievable schemes that achieve the best
known upperbound. For example, for ${5N}/{3} \leq M < 2 N$ half of the
streams are sent in the nullspace of one receiver and are aligned
with other interference streams at the other receiver. However, as
the ratio between $M$ and $N$ get smaller, a smaller number of
streams can be directed towards the nullspace and interference
alignment schemes are needed for more streams.

\begin{figure*}[!b]
\normalsize
\setcounter{mytempeqncnt}{\value{equation}}
\setcounter{equation}{8} \hrulefill
\begin{eqnarray}
{\boldsymbol{V}}_{\langle i\rangle_3}^{(j_{\langle i\rangle_3})}\!\!\!\!\!&\in&\!\!\!\!\! \mathcal{N}\{\boldsymbol{H}_{{\langle i+1\rangle_3},{\langle i \rangle}_3}\}\label{equi11a}\\
\boldsymbol{H}_{{\langle i+2 \rangle}_3,{\langle i+1\rangle}_3} {\boldsymbol{V}}_{{\langle i+1\rangle}_3}^{(j_{\langle i+1\rangle_3})}\!\!\!\!&=&\!\!\!\!\boldsymbol{H}_{{\langle i+2\rangle}_3,{\langle i\rangle}_3}{\boldsymbol{V}}_{{\langle i\rangle}_3}^{(j_{\langle i\rangle_3})}\label{equi13aa}\\
\boldsymbol{H}_{{\langle i \rangle}_3,{\langle i+2\rangle}_3} {\boldsymbol{V}}_{{\langle i+2\rangle}_3}^{(j_{\langle i+2\rangle_3})}\!\!\!\!&=&\!\!\!\!\boldsymbol{H}_{{\langle i\rangle}_3,{\langle i+1\rangle}_3}{\boldsymbol{V}}_{{\langle i+1\rangle}_3}^{(j_{\langle i+1\rangle_3})}\label{equi14}\\
\boldsymbol{H}_{{\langle i+1 \rangle}_3,{\langle i+3\rangle}_3} {\boldsymbol{V}}_{{\langle i+3\rangle}_3}^{(j_{\langle i\rangle_3}+1)}\!\!\!\!&=&\!\!\!\!\boldsymbol{H}_{{\langle i+1\rangle}_3,{\langle i+2\rangle}_3}{\boldsymbol{V}}_{{\langle i+2\rangle}_3}^{(j_{\langle i+2\rangle_3})}\label{equi15}\\
\boldsymbol{H}_{{\langle i+5 \rangle}_3,{\langle i+4 \rangle}_3} {\boldsymbol{V}}_{{\langle i+4\rangle}_3}^{(j_{\langle i+1\rangle_3}+1)}\!\!\!\!&=&\!\!\!\!\boldsymbol{H}_{{\langle i+5 \rangle}_3,{\langle i+3\rangle}_3}{\boldsymbol{V}}_{{\langle i+3\rangle}_3}^{(j_{\langle i\rangle_3}+1)}\label{equi15}\\
           &\vdots&\nonumber\\
\boldsymbol{H}_{{\langle L+i\rangle}_3,{\langle L+i-1\rangle}_3} {\boldsymbol{V}}_{{\langle L+i-1\rangle)}_3}^{( j_{\langle L+i-1\rangle_3}+  \lfloor\frac{L-1}{3} \rfloor       )}\!\!\!\!&=&\!\!\!\!\boldsymbol{H}_{{\langle L+i\rangle}_3,{\langle L+i-2\rangle}_3}{\boldsymbol{V}}_{{\langle L+i-2\rangle}_3}^{( j_{\langle L+i-2\rangle_3}+   \lfloor\frac{L-2}{3} \rfloor    )           }\;\label{equi16}\\
\boldsymbol{H}_{{\langle L+i-2\rangle}_3,{\langle L+i\rangle}_3} {\boldsymbol{V}}_{{\langle L+i\rangle}_3}^{ ( j_{\langle L+i\rangle_3}+   \lfloor\frac{L}{3} \rfloor       )}\!\!\!\!&=&\!\!\!\!\boldsymbol{H}_{{\langle L+i-2\rangle}_3,{\langle L+i-1\rangle}_3}{\boldsymbol{V}}_{{\langle L+i-1\rangle}_3}^{( j_{\langle L+i-1\rangle_3}+   \lfloor\frac{L-1}{3} \rfloor   )}\;\label{equi17}\\
{\boldsymbol{V}}_{{\langle L+i\rangle}_3}^{( j_{\langle
L+i\rangle_3}+\lfloor \frac{L}{3}\rfloor )}\!\!\!\!&\in&\!\!\!\!
\mathcal{N}\{\boldsymbol{H}_{{\langle L+i-1\rangle}_3,{\langle
L+i\rangle}_3}\}\;\label{equi12a}
\end{eqnarray}
\setcounter{equation}{\value{mytempeqncnt}}

\vspace*{4pt}
\end{figure*}

We will present the proposed achievable schemes for different
cases of the ratio $M/N$.

\subsubsection{$M/N$ larger than or equal to 3}

In this case, the number of DoF is upperbounded by $3N$.
Furthermore, the zero forcing scheme of \cite{Jafar_MN} can
achieve the optimal number of DoF.

\subsubsection{$M/N$ larger than or equal to 2 and less than 3}

In this case, it can be easily shown from (\ref{upperbound1_1})
that the number of DoF is upperbounded by $M$. The upperbound can
be achieved by assigning each user $d=M/3$ DoF via choosing the
precoding matrix of the $i$th user as
\begin{eqnarray}
\boldsymbol{V}_i&\!\!\!=\!\!\!&\mathcal{N}\{\boldsymbol{U}_j^H\boldsymbol{H}_{j,i}\}\bigcap \mathcal{N}\{\boldsymbol{U}_l^H\boldsymbol{H}_{l,i}\}\\
&\!\!\!=\!\!\!& \mathcal{N}\left\{\left[%
\begin{array}{c}
  \boldsymbol{U}_j^H\boldsymbol{H}_{j,i} \\
  \boldsymbol{U}_l^H\boldsymbol{H}_{l,i} \\
\end{array}%
\right] \right\} \label{null_intersection}
\end{eqnarray}
where $\mathcal{N}\{\boldsymbol{A}\}$ denotes the nullspace of the
matrix $\boldsymbol{A}$, the indices $(l,j,i)=(1,2,3), (2,3,1),
(3,1,2)$, and the decoding matrices $\{\boldsymbol{U}_i\}_{i=1}^3$
are selected randomly. Note that the dimension of the subspace in
(\ref{null_intersection}) is given by $M-2d$, and hence, $d=M/3$
streams can be transmitted from each user.

\subsubsection{$M/N$ smaller than 2}

We here present the main scheme of the paper. Our scheme starts by
dividing the precoding matrix of each transmitter into a number of
parts where each part performs interference alignment
independently. The alignment conditions of these parts are
different, some parts totally align their signal such that they
lie in the interference subspace at both non-intended receivers,
others align their signal to lie in the interference subspace of
one receiver and completely direct their interference in the
nullspace of the other non-intended receiver.

Let us divide the $M\times d_i$-dimensional precoding matrix for
the $i$th transmitter into $L+1$ subblocks each of size $M\times
\tilde{d}$
\begin{equation}\label{V_i_partition}
    \boldsymbol{V}_i=\left[\boldsymbol{V}_i^{(1)}, \boldsymbol{V}_i^{(2)}, \ldots,
    \boldsymbol{V}_i^{(L+1)}\right].
\end{equation}
Hence, the total number of streams transmitted by the $i$th user
is given by $d_i= \tilde{d}(L+1) $ and the total number of DoF is
given by $d= 3\tilde{d}(L+1)$. Our scheme can be described by
equations (\ref{equi11a})--(\ref{equi12a}) at the bottom of next
page where ${{\langle \cdot \rangle}_n}$ denotes the modulo-$n$
operator, $\langle n \rangle_n=n$, and $\lfloor x \rfloor$ denotes
the largest integer that is less than or equal to $x$. The system
of equations in (\ref{equi11a})--(\ref{equi12a}) is repeated three
times where in the first time $(i,j_1,j_2,j_3)=(1,1,1,1)$. Note
that $j_k-1$ indicates the number of partitions of the matrix
$\boldsymbol{V}_k$ that has appeared in the previous, i.e.,
$(i-1)$th, set of equations. Hence, for the second set, we have
$i=2$ and $j_k=\lceil(L+1)/3\rceil + 1$ if $k \leq \langle
L+1\rangle_3$ and $\lfloor(L+1)/3\rfloor+1$ else. Also, for the
third set, we have $i=3$ and $j_k$ is incremented by
$\lceil(L+1)/3\rceil$ if $\langle k-1 \rangle_3 \leq \langle
L+1\rangle_3$ and by $\lfloor(L+1)/3\rfloor$ else.

\addtocounter{equation}{8}

Note that the parameter $L$ completely specifies the above system
of equations. The number of equations in
(\ref{equi11a})--(\ref{equi12a}) is given by $L+2$. Equation
(\ref{equi11a}) constrains the transmission of the matrix
${\boldsymbol{V}}_{\langle i\rangle_3}^{(j_{\langle i\rangle_3})}$
to lie in the nullspace of $\boldsymbol{H}_{{\langle
i+1\rangle_3},{\langle i \rangle}_3}$, and hence, it does not
cause any interference on the ${\langle i+1\rangle_3}$th receiver.
The next $L$ equations align the interference from the precoding
submatrices of each two transmitters at the interference subspace
of the third receiver, for example, equation (\ref{equi13aa})
aligns the interference caused by the transmission of
${\boldsymbol{V}}_{{\langle i+1\rangle}_3}^{(j_{\langle
i+1\rangle_3})}$ at the ${\langle i+2\rangle}_3$th receiver with
the interference caused by the transmission of
${\boldsymbol{V}}_{\langle i\rangle_3}^{(j_{\langle i\rangle_3})}$
at the same receiver. The final equation constrains the
transmission of the matrix ${\boldsymbol{V}}_{{\langle
L+i\rangle}_3}^{( j_{\langle L+i\rangle_3}+\lfloor
\frac{L}{3}\rfloor )}$ to lie in the nullspace of
$\boldsymbol{H}_{{\langle L+i-1\rangle}_3,{\langle L+i\rangle}_3}$,
and hence, it does not cause any interference on the ${\langle
L+i-1\rangle}_3$th receiver. Note that the system of equations in
(\ref{equi11a})--(\ref{equi12a}) constrains the transmission of
$L+1$ subblocks of the precoding matrices of the three users that
are transmitting a total number of $(L+1) \tilde{d}$ streams in
the signal subspaces of the $3$ receivers. Furthermore, these
transmissions are constrained to lie in an interference subspace
of dimension $L\tilde{d}$ distributed among the three receivers.
Hence, the total number of dimensions consumed at the $3$
receivers (interference+signal) by the complete system of
equations corresponding to $i=1,2,3$ is given by
$3(2L+1)\tilde{d}$. Since the total number of receive dimensions
is given by $3N$, and the total number of streams transmitted by
the $i$th user is given by $d_i= \tilde{d}(L+1)$, we have the
following upperbound on the total number of achievable DoF of our
scheme
\begin{equation}\label{limit_on_N}
{d} \leq \frac{3L+3}{2L+1} N.
\end{equation}
Later in this section, we will show the relationship between the
parameter $L$ and the number of transmit and receive antennas such
that there is no interference at any receiver.

Before explaining how to obtain a solution of the above system of
equations, and presenting the conditions on the number of
achievable DoF, we present the case which has only one alignment
equation, i.e., $L=1$. In this case, the complete system of
equations is given by
\vspace{-1mm}
\begin{eqnarray}
\boldsymbol{V}_1^{(1)}&\in& \mathcal{N}\{\boldsymbol{H}_{2,1}\}\label{equi11acc}\\
\boldsymbol{H}_{3,2}\boldsymbol{V}_2^{(1)}&= &\boldsymbol{H}_{3,1}\boldsymbol{V}_1^{(1)}\label{equi13acc}\\
\boldsymbol{V}_2^{(1)}&\in& \mathcal{N}\{\boldsymbol{H}_{1,2}\}\label{equi12acc}\\
\boldsymbol{V}_2^{(2)}&\in& \mathcal{N}\{\boldsymbol{H}_{3,2}\}\label{equi11bcc}\\
\boldsymbol{H}_{1,3}\boldsymbol{V}_3^{(1)}&= &\boldsymbol{H}_{1,2}\boldsymbol{V}_2^{(2)}\label{equi13bcc}\\
\boldsymbol{V}_3^{(1)}&\in& \mathcal{N}\{\boldsymbol{H}_{2,3}\}\label{equi12bcc}\\
\boldsymbol{V}_3^{(2)}&\in& \mathcal{N}\{\boldsymbol{H}_{1,3}\}\label{equi11ccc}\\
\boldsymbol{H}_{2,1}\boldsymbol{V}_1^{(2)}&= &\boldsymbol{H}_{2,3}\boldsymbol{V}_3^{(2)}\label{equi13ccc}\\
\boldsymbol{V}_1^{(2)}&\in&
\mathcal{N}\{\boldsymbol{H}_{3,1}\}\label{equi12ccc}
\end{eqnarray}
We will show how we can find $\boldsymbol{V}_1$,
$\boldsymbol{V}_2$, and $\boldsymbol{V}_3$ that satisfy the system
of equations, and find the number of achievable DoF. Let
$\boldsymbol{\Xi}_{i,j}$ denote the $M \times M-N$ matrix whose
columns span nullspace of the matrix $\boldsymbol{H}_{i,j}$. Let
us consider the first group of equations, i.e., the first $L+2$
equations. From (\ref{equi11acc}), (\ref{equi13acc}) and
(\ref{equi12acc}) respectively, we can write
\begin{eqnarray}
\vspace{-1mm}
\boldsymbol{V}_1^{(1)} &=& \boldsymbol{\Xi}_{2,1} \boldsymbol{A}_{2,1} \label{eq1}\\
\boldsymbol{V}_2^{(1)} &=& \boldsymbol{H}_{3,2}^{\dag}
\boldsymbol{H}_{3,1}\boldsymbol{V}_1^{(1)} +
\boldsymbol{\Xi}_{3,2}
\boldsymbol{A}_{3,2} \label{eq2}\\
\boldsymbol{V}_2^{(1)} &=& \boldsymbol{\Xi}_{1,2}
\boldsymbol{A}_{1,2} \label{eq3}
\end{eqnarray}
where $\boldsymbol{A}^{\dag}$ denotes the pseudo-inverse of the
matrix $\boldsymbol{A}$ and the dimensions of the matrices
$\boldsymbol{A}_{2,1}$, $\boldsymbol{A}_{3,2}$, and
$\boldsymbol{A}_{2,1}$ are $M-N \times \tilde{d}$. Substituting
with (\ref{eq1}), and (\ref{eq3}) in (\ref{eq2}), we get
\begin{equation}
\boldsymbol{\Xi}_{1,2} \boldsymbol{A}_{1,2}=
\boldsymbol{H}_{3,2}^{\dag} \boldsymbol{H}_{3,1}
\boldsymbol{\Xi}_{2,1} \boldsymbol{A}_{2,1} +
\boldsymbol{\Xi}_{3,2} \boldsymbol{A}_{3,2}
\end{equation}
and hence, we can write
\begin{equation}
\tilde{{\boldsymbol{\Xi}}}_1 \tilde{\boldsymbol{A}}_1=
\boldsymbol{0}_{M \times \tilde{d}}
\end{equation}
where the $3(M-N)\times \tilde{d}$ matrix $\tilde{\boldsymbol{A}}=
\left(\boldsymbol{A}_{1,2}^T, \boldsymbol{A}_{2,1}^T,
\boldsymbol{A}_{3,2}^T \right)^T$ and the ${M\times 3 (M-N)}$
matrix $\tilde{\boldsymbol{\Xi}}_1$ is given by
\begin{equation}
\tilde{\boldsymbol{\Xi}}_1= \left(-\boldsymbol{\Xi}_{1,2},
\boldsymbol{H}_{3,2}^{\dag} \boldsymbol{H}_{3,1}
\boldsymbol{\Xi}_{2,1}, \boldsymbol{\Xi}_{3,2} \right)
\end{equation}
Since the dimension of the nullspace of
$\tilde{{\boldsymbol{\Xi}}}_1$ is given by $2M-3N$, we obtain the
following bound on the number of streams
\begin{equation}\label{bound_d_tilde1}
\tilde{d} \leq \left(2M-3N\right)^{+}
\end{equation}
where $(x)^{+}$ denotes $\max\{x,0\}$.

Given the nullspace of the matrix $\tilde{{\boldsymbol{\Xi}}}_1$,
we can compute the precoding matrices $\boldsymbol{V}_1^{(1)}$ and
$\boldsymbol{V}_2^{(1)}$ using (\ref{eq1}), and (\ref{eq3}). The
same procedure can be applied on equations
(\ref{equi11bcc})--(\ref{equi12bcc}) to evaluate the matrices
$\boldsymbol{V}_2^{(2)}$ and $\boldsymbol{V}_3^{(1)}$ and on
equations (\ref{equi11ccc})--(\ref{equi12ccc}) to evaluate the
matrices $\boldsymbol{V}_1^{(2)}$ and $\boldsymbol{V}_3^{(2)}$.
Hence, the total number of  DoF that can be achieved in the
network is upperbounded by
\begin{equation}\label{bound_d_tilde1}
d \leq (12M-18N)^{+}
\end{equation}
In addition, we have the upperbound on the achieved DoF by the
proposed scheme in (\ref{limit_on_N}). Therefore, the total number
of achieved DoF obtained by solving the sets of equations for
$L=1$ is upperbounded by
\begin{equation}\label{bound_d_tilde2}
d \leq \min\{(12M-18N)^{+}, 2N\}.
\end{equation}

For $5/3 \leq M/N < 2$, the general upperbound on the DoF in
(\ref{upperbound1_1}) is equal to $2N$, and is tighter than the
beamforming upperbound in (\ref{upperbound2}). In this region, the
number of DoF obtained by solving the sets of equations for $L=1$
is given by $2N$ since $\min\{(12M-18N)^{+}, 2N\}=2N$. Hence, the
proposed scheme with $L=1$ can achieve the maximum number of DoF
available in the network and therefore is DoF-optimal. Also, note
that at $M/N=5/3$, the two general and the beamforming upperbounds
in (\ref{upperbound1_1}) and (\ref{upperbound2}) are equal and are
achieved by the proposed scheme with $L=1$.

\begin{figure}[]
  \begin{center}\vspace{-.5mm}
 \includegraphics[width=.450\textwidth]{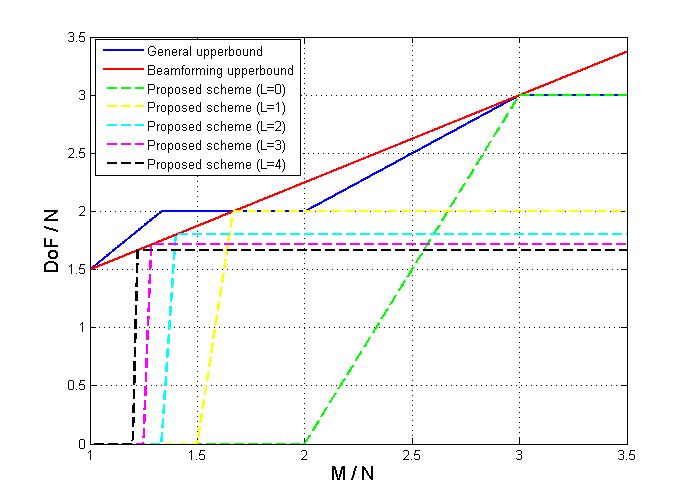}\label{graph}\vspace{-2mm}
 \caption{Number of achievable DoF by the proposed scheme for different $L$}
  \label{fig one}
  \end{center} \vspace{-7.5mm}
  \end{figure}

Let us return to the general scheme with $3 (L+2)$ equations in
(\ref{equi11a})--(\ref{equi12a}). We can write \small
\begin{eqnarray}
{\boldsymbol{V}}_{\langle i\rangle_3}^{(j_{\langle i\rangle_3})}\!\!\!\!\!\!&=&\!\!\!\!\!\boldsymbol{\Xi}_{{\langle i+1\rangle_3},{\langle i\rangle}_3} \boldsymbol{A}_{{\langle i+1\rangle},{\langle i\rangle}}  \label{equi11bb}\\
 {\boldsymbol{V}}_{{\langle i+1\rangle}_3}^{(j_{\langle i+1\rangle_3})}\!\!\!\!\!\!&=&\!\!\!\!\! \boldsymbol{\Xi}_{{\langle i+2\rangle}_3,{\langle i+1\rangle}_3}
\boldsymbol{A}_{{\langle i+2 \rangle},{\langle i+1\rangle}}
\nonumber\\
\!\!\!\!\!\!&+&\!\!\!\!\! \boldsymbol{H}_{{\langle i+2\rangle}_3,{\langle i+1\rangle}_3}^{\dag} \boldsymbol{H}_{{\langle i+2 \rangle}_3,{\langle i \rangle}_3}{\boldsymbol{V}}_{\langle i \rangle_3}^{(j_{\langle i\rangle_3})} \label{equi12bb}\\
           &\vdots&\nonumber\\
\boldsymbol{V}_{{\langle L+i\rangle}_3}^{(j_{\langle
L+i\rangle_3}+\lfloor \frac{L}{3}\rfloor
)}\!\!\!\!\!\!&=&\!\!\!\!\!\boldsymbol{\Xi}_{{\langle
L+i-2\rangle}_3,{\langle L+i\rangle}_3} \boldsymbol{A}_{{\langle
L+i-2\rangle},{\langle L+i\rangle}}   \qquad\nonumber\\
\!\!\!\!\!\!&&\!\!\!\!\! \hspace{-2.3cm}+\boldsymbol{H}_{{\langle
L+i-2\rangle}_3,{\langle L+i\rangle}_3}^{\dag}
\boldsymbol{H}_{{\langle L+i-2\rangle}_3,{\langle L+i-1\rangle}_3}
\boldsymbol{V}_{{\langle L+i-1\rangle}_3}^{( j_{\langle
L+i-1\rangle_3}+
\lfloor\frac{L-1}{3} \rfloor   )} \quad\label{equi13bb}\\
\boldsymbol{V}_{{\langle L+i\rangle}_3}^{(j_{\langle
L+i\rangle_3}+\lfloor \frac{L}{3}\rfloor
)}\!\!\!\!\!\!&=&\!\!\!\!\! \boldsymbol{\Xi}_{{\langle
L+i-1\rangle}_3,{\langle L+i\rangle}_3} \boldsymbol{A}_{{\langle
L+i-1\rangle},{\langle L+i\rangle}}\label{equi14bb}
\end{eqnarray} \normalsize
Therefore, we can write
\begin{equation}
    \tilde{\boldsymbol{\Xi}}_i \tilde{\boldsymbol{A}}_i =
    \boldsymbol{0}_{M \times \tilde{d}}
\end{equation}
where $i=1,2,3$ and the dimensions of the matrices
$\tilde{\boldsymbol{\Xi}}_i$ and $\tilde{\boldsymbol{A}}_i$ are
given by $M\times(L+2)(M-N)$ and $(L+2)(M-N)\times\tilde{d}$,
respectively. Since the dimension of the nullspace of
$\tilde{\boldsymbol{\Xi}}_i$ is $(L+1)M-(L+2)N$, we can achieve
the following DoF
\begin{equation}
d \leq 3 (L+1) \left((L+1)M-(L+2)N\right)^{+}.
\end{equation}
Combining the above upperbound on the achievable DoF by the
proposed scheme with that in (\ref{limit_on_N}), we get
\begin{equation}\label{bound_d_tilde_general_L}
d \leq \min\{ 3 (L+1) \left((L+1)M-(L+2)N\right)^{+},
\frac{3L+3}{2L+1} N \}.
\end{equation}
Note that when ${M}/{N} \leq {(2L+3)}/{(2L+1)}$, we have
$\left((L+1)M-(L+2)N\right)^{+} \leq N /(2L+1)$, and hence, the
total number of DoF achieved by the proposed scheme is given by
\begin{equation}
d = 3 (L+1) \left((L+1)M-(L+2)N\right)^{+}.
\end{equation}
Furthermore, at ${M}/{N} = {(2L+3)}/{(2L+1)}$, the number of
achieved DoF meets the beamforming upperbound in
(\ref{upperbound2}) which is tighter than the general upperbound
in (\ref{upperbound1_1}) in this case. Hence, the proposed scheme achieves the maximum
DoF available through beamforming when ${M}/{N} = {(2L+3)}/{(2L+1)}$. Fig.~\ref{fig one}
shows the number of achievable DoF by the proposed scheme at
different values of the parameter $L$ versus the ratio $M/N$ and
their relationship to the general and beamforming upperbounds.

\subsection{Combinatorics}
We have see from the previous subsection that if ${M}/{N} \neq
{(2L+3)}/{(2L+1)}$ for some integer $L$, the proposed scheme is
not DoF-optimal. Nevertheless, we can design our precoding
matrices using more than one value of $L$ to achieve more DoF than
those achieved by using only one value of $L$. Note that we can
utilize any value of $L$ as long as $(L+1)M-(L+2)N$ is larger than
or equal to $1$, so that we can find a solution to the system of
equations in (\ref{equi11a})--(\ref{equi12a}). We also notice
that for each $L+1$ streams sent in the network we consume $2L+1$
receiver dimensions. It is hence more efficient to use smaller
values of $L$. However, the streams sent using a certain $L$ might
not ``fill" all the receiver dimensions. Hence, an efficient
scheme would start with the lowest possible value of $L$, send the
highest possible number of streams using this value. If there is unutilized receiver dimensions, we design precoding matrices
for extra streams using the next value of $L$ and so on.

In order to illustrate this design technique, Let us assume that
$M=30$ and $N=19$. Setting $L=1$, we can design
$d_i=(L+1)(2M-3N)=6$ streams per user while consuming
$(2L+1)(2M-3N)=9$ receiver dimensions per user. Since $N=19$, we
still have $19-9=10$ receive dimensions per user to exploit. Hence
we then use $L=2$. Since, $3M-4N=14$, we can design up to
$14(L+1)$ streams per user, but we will be limited by the receiver
dimensions. For $L=2$, each $(L+1)=3$ streams sent per user
utilize $(2L+1)=5$ receiver dimensions. Hence, we can send $6$
more streams per user bringing the total number of streams to $12$
streams per user or $36$ streams for the overall system which is
close to the beamforming upperbound of $3(M+N)/4=36.75$.

We have assumed throughout this paper that all users use the same
number of streams. It can be actually shown that in some cases,
different users should use different number of streams. To find
the optimal solution, i.e. one that maximizes the total number of
system streams, the problem can be posed as an integer programming
problem. We will explore this in an extended version of this
paper.

\vspace{-2mm}

\section{Numerical results}

\begin{figure}[]
  \begin{center}\vspace{-.5mm}
 \includegraphics[width=.450\textwidth]{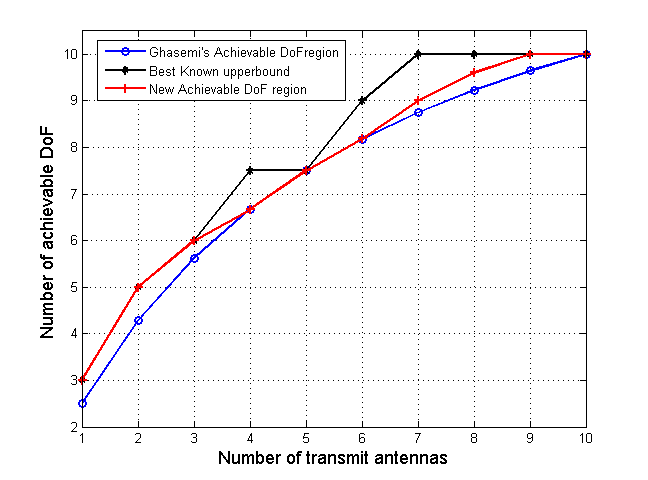}\label{graph}\vspace{-2mm}
  \caption{Number of DoF that can be obtained by new schemes}
  \label{fig two} \vspace{-8.5mm}
  \end{center}\end{figure}

In order to illustrate what our scheme really add, we give
the total number of achievable DoF for $N=5$ and $M=1,2,...,16$.
Fig.~\ref{fig two} compares the new DoF region we obtained to the
best known DoF region found \cite{Khandani_KuserMxN} and to the
best known upperbound. For $N > M$ the number of DoF are equal to
the DoF of the reverse link and Fig.~\ref{fig two}  shows the DoF
with number of antennas at both the transmitter and the receiver
exchanged. The fractional DoF is obtained using finite symbol
extension in time. For $M=4$, $N=5$, the scheme is used with $L=4$
over $9$ time slots. This scheme wastes $4$ dimensions for
interference for every $5$ streams sent, a total of $9$
dimensions needed at the receiver for every $5$ DoF, so $9$ time
slots is used to make the total number of resources a multiple of
9 and maximize the total number of DoF. For $M=8$,
the scheme is used twice with $L=1,2$ over $5$ time slots. First
the scheme is used with $L=1$ to obtain $2$ DoF per user per time slot,
then it is used with $L=2$ to obtain $1.2$ DoF per user per time slot.  We
achieve more DoF than known before for $M=2,3,7,8,9$. Our region
actually meets the upperbound at $M=2,3,9$. This leaves the number
of DoF for $M=4,6,7,8$ unknown. 

  \vspace{-3mm}
\section{Conclusion}
We have provided a new achievable interference alignment scheme
for $3$-user MIMO Gaussian interference channel with constant
channel coefficients. We have showed that the proposed scheme
spans a new achievable DoF region. We have also showed that for
some values of $M$ and $N$ it meets the best known upperbound.

\vspace{-3mm}
\bibliographystyle{IEEEbib}
\bibliography{Thirdv8}
\end{document}